\documentclass[authoryear]{elsarticle}

\usepackage{rotating}
\usepackage{graphics}
\usepackage{graphicx}
\usepackage{epstopdf}
\usepackage{color}
\usepackage{amssymb}
\journal{Semicond. and Semimetals 85, 47 (2011)}
\def\re#1{(\ref{#1})}\bibliography{}
\def\be{\begin{equation}}
\def\ee{\end{equation}}
\def\bea{\begin{eqnarray}}
\def\eea{\end{eqnarray}}
\def\bra#1{ \langle #1 |}
\def\ket#1{|#1\rangle}

\def\cm{cm$^{-1}\,$}
\begin{document}
\begin{frontmatter}
\title{Quantum Dynamics and Spectroscopy of Excitons in Molecular Aggregates}
\author[UR]{Oliver K\"uhn\corref{cor1}}
\cortext[cor1]{Corresponding author}
\ead{oliver.kuehn@uni-rostock.de}
\author[UR]{Stefan Lochbrunner}

\address[UR]{Institut f\"ur Physik, Universit\"at Rostock, D-18051 Rostock, Germany}

\begin{abstract}
The theoretical description and the properties of Frenkel excitons in non-cova\-lent\-ly bonded molecular aggregates are reviewed from a multi-exciton perspective of dissipative quantum dynamics. First, the photophysical and quantum chemical characterization of the monomeric dye building blocks is discussed, including the important aspect of electron-vibrational coupling within the Huang-Rhys model. Supplementing the model by the Coulombic interactions between monomers, the description of aggregates in terms of excitonic or vibrational-excitonic bands follows. Besides of giving rise to complex absorption and emission line shapes, exciton-vibrational interaction is responsible for energy and phase relaxation and thereby limits the size of coherent excitations in larger aggregates. Throughout, emphasis is put on the electronic three-level model as a minimum requirement to describe nonlinear spectroscopies including effects of two-exciton states such as excited state absorption and exciton-exciton annihilation. The experimentally observed characteristics of stationary absorption and fluorescence spectra of aggregates as well as their temperature dependence are discussed. Examples for ultrafast spectroscopic experiments including pump-probe studies, photon echo and two-dimensional spectroscopy are presented and results on the size of coherence domains and on intra- and interband relaxation are given. Finally, experimental signatures for exciton-exciton annihilation and their analysis with respect to the mobility of excitons are described.
\end{abstract}

\end{frontmatter}

%
\section{Introduction}
\label{sec:intro}
%
Organic solar cells are a highly active research field since they hold the promise to provide a cheap source for renewable energy. In such cells the conversion process of a photon into electric energy consists of several steps (\cite{Bra01,gregg03:4688a,Ayz09}). First the photon is absorbed by dye molecules or a polymer and an exciton is generated. The exciton has to reach an interface between two materials with shifted band gaps. Here charge separation can take place. The generated charges have then to migrate to the contacts. In commonly used organic materials the exciton diffusion length is limited to about 10~nm (\cite{Scu06,Mik08}). This short distance poses severe restrictions to the design of organic solar cells. A wide spread concept to deal with the short exciton diffusion length are bulk heterojunctions in which two materials form strongly interleaving domains (\cite{Bra01}). However, not too many materials are suitable for bulk heterojunctions. Another option is to design organic materials with a high degree of order which provide a long exciton diffusion length. For this purpose aggregates are highly attractive since, in addition to high exciton mobilities, they exhibit also strong absorption bands in the visible spectral region (\cite{mobius95:437, wuerthner11:3376}). Aggregates are highly ordered supramolecular structures of dye molecules which form by self-assembling. The monomers are bound by van-der-Waals forces and often in addition by hydrogen bonds. To understand and design their excitonic properties is crucial for their application in organic solar cells as well as in other photonic devices. The theoretical description and spectroscopic investigation of these excitonic properties are the topic of this review.

Molecular aggregates store and transfer electronic exciton energy by means of  Frenkel excitons (\cite{knoester03:1,beljonne09:6583,may11}). 
Frenkel excitons are electron-hole pairs localized at the same molecular unit (monomer) of an aggregate. They have to be distinguished from Wannier-Mott excitons common in solids, where the electron-hole separation reaches hundreds of nanometers. Further, so-called charge transfer excitons play a role in molecular systems, provided that a means for charge separation e.g. between parts of a larger molecular unit is given.  Exciton formation has been observed first by Jelly and Scheibe in 1936, who detected a narrowing of the relatively broad absorption band of pseudo-isocyanine (PIC) upon increasing its concentration in solution (for reviews, see \cite{mobius95:437} and \cite{wuerthner11:3376}).
Nowadays it is recognized that the concept of excitons is not restricted to molecular aggregates, but is a basic mechanism for electronic excitation energy transfer (EET) in various nanoscale and biological systems (\cite{scholes06:683}).

Traditionally the theoretical description of electronic EET  has been based on the model of electronic two-level systems, coupled via dipole-dipole interactions  (\cite{foerster48:55,kasha65:371}). Considering the aggregate to be build up of monomers which retain their chemical identity, the dipole-dipole coupling is an approximation to the full Coulomb interaction between the electron densities of the monomers. Its limitation is obvious and especially severe in case of tightly packed aggregates. In recent years the advances in computational power and numerical methods have made it possible to determine the full Coulomb coupling between monomers from first principles, even for rather large dye molecules.

\begin{figure}
\begin{center}
\includegraphics[width=0.65\textwidth]{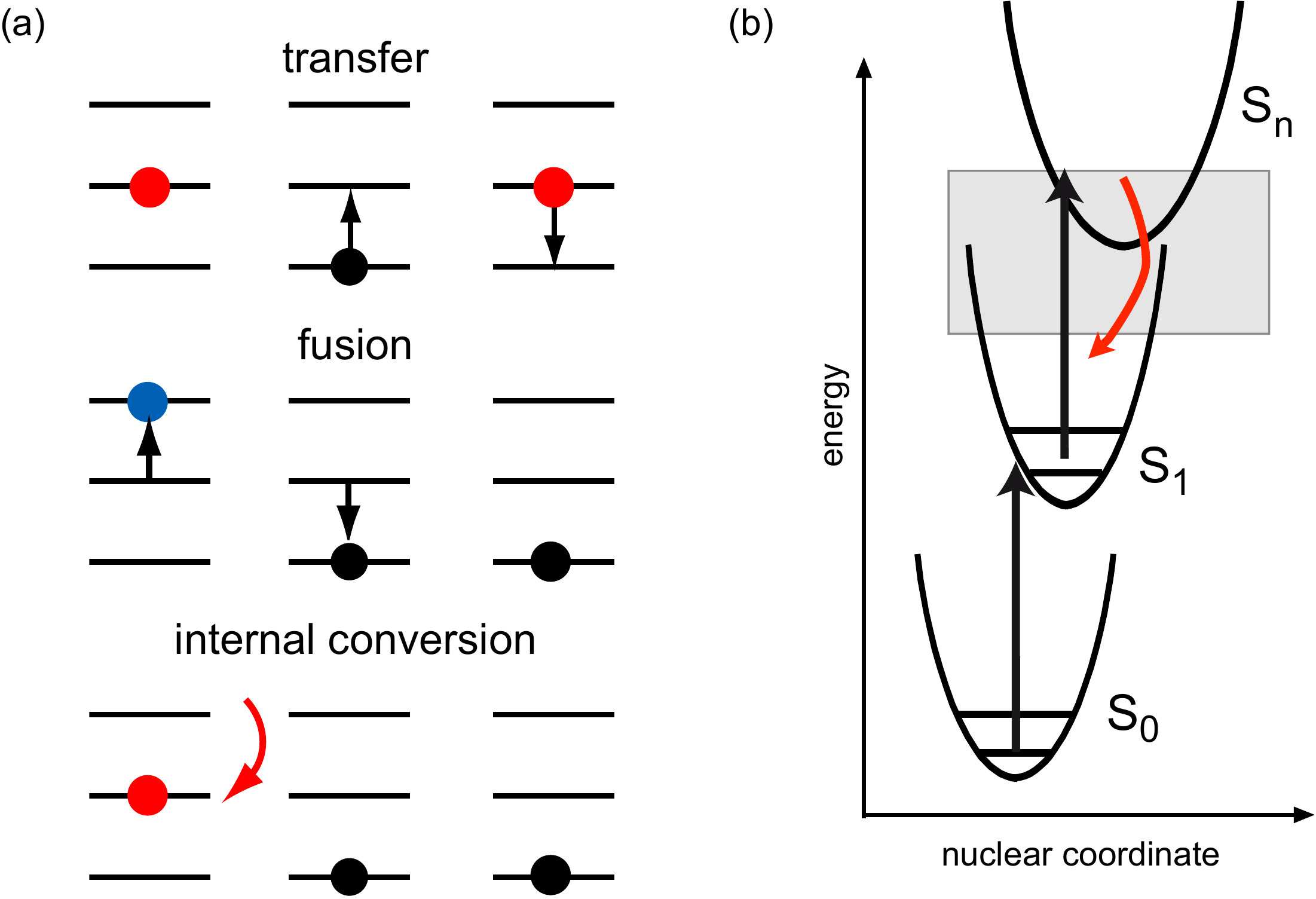}
\caption{(a) Different processes in aggregates composed of three level monomers. (b) Excitation and internal conversion process in a model of displaced harmonic oscillators describing the nuclear motions of the monomer. Notice that the internal conversion process usually proceeds via intermediate electronic states as indicated by the shaded area.}
\label{fig:scheme}
\end{center}
\end{figure}
Scrutinizing the validity of the two-level approximation is equally important. In fact for the understanding of linear spectroscopies this model is well suited. However, for the description of nonlinear spectroscopy as well as of processes triggered by  higher laser intensities the incorporation of higher excited electronic states might become vital. Here, more  photons are absorbed, usually at similar energies. Viewed from the monomeric perspective of Fig. \ref{fig:scheme}b, after, e.g. a S$_0 \rightarrow$ S$_1$ excitation, a second photon may excite a  S$_1 \rightarrow$ S$_n$ transition, where $n$ could be of the order of 10 or higher. Thus, strictly speaking the modeling of all processes following this excited state absorption (ESA) would require the use of $n$ electronic levels for each monomer. However, according to Kasha's rule it can be expected that the S$_n$ state population quickly relaxes down to the S$_1$ state. Thus to a good approximation the monomer can be described by an electronic three-level model, supplemented by a rapid internal conversion mechanism between the S$_n$ and S$_1$ states. 

There is a second type of excitation, which can be reached upon absorption of two photons, i.e. two monomers might be excited to the S$_1$ state yielding a nonlocal doubly excited (two-exciton) state of the aggregate. Besides the importance of these local and nonlocal doubly excited states for spectroscopy, their interplay has some consequences for the dynamics of EET as illustrated in Fig. \ref{fig:scheme}a. Consider a nonlocal double excitation (upper panel). The two excitations can transfer independently through the system, but  owing to the Pauli principle a monomer S$_1$ state cannot be doubly excited. However, the two excitations can combine to yield a local S$_n$ excitation (exciton fusion, middle panel). The life time of the S$_n$ state, however, is usually short and internal conversion will quickly lead the system back to the S$_1$ state. Hence, in total one electronic excitation has been dissipated into heat, i.e. vibrational motions. This so-called exciton-exciton annihilation (EEA) mechanism is an important limiting factor for device applications as it restricts the amount of excitation which can be taken up and transported by the aggregate. 

The microscopic origin of EEA is the breakdown of the Born-Oppenheimer approximation, leading to non-adiabatic couplings between the adiabatic electronic states triggered by nuclear motions. However, even within the Born-Oppenheimer approximation there will be effects of electron-vibrational coupling on the dynamics and spectroscopy of Frenkel excitons. This includes the appearance of vibrational progressions in particular for isolated higher frequency vibrational modes in the absorption and emission spectra, which combined with the Coulomb coupling may result in rather complex line shapes. The coupling to the dense manifold of low-frequency vibrational modes will cause an overall line broadening, where one can distinguish between static and dynamic disorder according to the time scale of nuclear motions with respect to relevant times of the exciton motion. In the time domain the effect of exciton-vibrational coupling (EVC) is to cause exciton coherence dephasing and population relaxation. Thus EVC limits the size over which a coherent excitation can be established within an aggregate and therefore is crucial for exciton transport. 

In this review we will discuss exciton dynamics and spectroscopy focussing on some of the points just raised, namely the three-level nature of the monomeric building blocks, exciton-exciton annihilation, and exciton-vibrational coupling. We will restrict the discussion to molecular aggregates; for an account on excitons in biological pigment-protein complexes and artificial nanoscale systems like polymers we refer the reader to other contributions in this book. Furthermore, we will not treat the issue of exciton dissociation and free charge carrier formation which is crucial for the function of excitonic organic solar cells (\cite{gregg03:4688a}); for a review on recent theoretical advances see \cite{burghardt09:183}. After a brief account on various aspects of Frenkel exciton theory in Section \ref{sec:theory}, spectroscopic applications will be discussed in Section \ref{sec:appl}, which highlight the unique properties of excitons in molecular aggregates.
%
\section{Theory}
\label{sec:theory}
\subsection{Molecular Aggregate Hamiltonian}
Separating the aggregate into its constituent monomers, which are coupled by the Coulomb interaction, the aggregate Hamiltonian becomes (\cite{renger01:137,may11})
\bea
\label{eq:Hagg}
H &=& \sum\limits_m \sum\limits_{a, b} (\delta_{ab} H_{m,a} +  {\Theta}_{m, a b})
\ket{\varphi_{m a}}\bra{\varphi_{m b}}  \nonumber \\
&& + \frac{1}{2} \sum\limits_{m, n}
\sum\limits_{a, b, c, d} J_{m n}(a b, c d)
\ket{\varphi_{m a}} \bra{\varphi_{m d}} \otimes
\ket{\varphi_{n b}} \bra{\varphi_{n c}}  \; .
\eea
Here, $\ket{\varphi_{ma}}$ denotes the $a$th adiabatic electronic state of the $m$th monomer, which depends on the nuclear coordinates $R_{m}=(\mathbf{ R}_{m,1},\mathbf{ R}_{m,2},\ldots)$ of the monomer, i.e. $\varphi_{ma}=\varphi_{ma}(R_{m})$. This dependence is found also for the matrix elements entering Eq. \re{eq:Hagg}. For the monomer contribution we have
\bea
\label{c9-h-mab}
H_{m,a} = H_{m,a}(R_m) = T_m + V_{m a}(R_m) \, .
\eea
Here, $T_m$ and $V_{m a}(R_m)$ denote the nuclear kinetic energy and potential energy surface, respectively. Further, in Eq. \re{eq:Hagg} ${\Theta}_{m, a b}$ is the operator  of non--adiabatic coupling between different adiabatic electronic states and it is assumed that ${\Theta}_{m, a a}=0$ .
The second sum in Eq. \re{eq:Hagg} contains the Coulomb coupling between monomers $m$ and $n$. It can be expressed in terms of transition densities 
\bea
\label{eq:td}
n^{(m)}_{a b}({\mathbf{ x}}) = \varrho^{(m)}_{a b}({\mathbf{ x}}) - \delta_{ab} 
\sum\limits_{\mu } e Z_{\mu}\delta(\mathbf{ x} - \mathbf{ R}_{m,\mu}) \; 
\eea
as follows
\bea
\label{eq:jmn}
J_{m n}(a b, c d) = \int\limits d^3 \mathbf{ x} d^3  \mathbf{ x}' \; 
\frac{n^{(m)}_{a d}({\mathbf x}) n^{(n)}_{b c}({\mathbf x}')}{|{\mathbf x} - {\mathbf x}'|} \; .
\eea
In Eq. \re{eq:td} $\varrho^{(m)}_{a b}({\mathbf{ x}}) $ is the electronic transition density for a transition between the adiabatic states $\ket{\varphi_{ma}}$  and $\ket{\varphi_{mb}}$ at monomer $m$ and the $Z_\mu$ are the atomic numbers.
Often the Coulomb coupling is treated within the dipole approximation assuming that the spatial extension of the transition density is small as compared with the distance between the monomers (\cite{may11}). This gives
\bea
\label{c9-v12dd}
J^{\rm dip}_{m n}(a b, c d) =
\frac{ {\mathbf d}_{m, a d} \cdot{\mathbf d}_{n, b c} }{| {\mathbf X}_{m n} |^3}
- 3 \frac{( {\mathbf X}_{m n} \cdot{\mathbf d}_{m, a d})
( {\mathbf X}_{m n}\cdot{\mathbf d}_{n, b c})}{| {\mathbf X}_{m n} |^5} \; .
\eea
Here, ${\mathbf X}_{m n}$ is the distance vector between the considered monomers (e.g. center of mass) and the ${\mathbf d}_{m, a b}$ are transition dipole moments.

Considering the influence of nuclear motions on the EET dynamics one has to distinguish between the effect on the intramolecular transition energies (via $V_{m a}(R_m)$) and on the Coulomb coupling. For aggregates in solution there will be an additional dependence of both quantities on the solvent coordinates. As discussed in \cite{may11} there are different models to cope with this situation (cf. Sec. \ref{sec:evc}). This includes the description in terms of combined local intramolecular vibrations and intermolecular as well as solvent normal modes, the use of global normal modes for the total aggregate plus solvent system, and the classical description of nuclear motions. While the former two cases have been in use for quite some time, explicitly accounting for the thermal fluctuations by virtue of molecular dynamics simulation of aggregates has become feasible only recently (see, e.g. \cite{zhu07:118,zhu08:154905}).
\subsection{Frenkel Exciton Hamiltonian for Coupled Three-Level Monomers}
EET is commonly discussed on the basis of an electronic two-level approximation for the monomers. While this is sufficient for the discussion of linear optical properties, the understanding of nonlinear optical experiments as well as of the dynamics at higher excitation densities requires to take into account the fact that organic dye molecules forming the aggregate show ESA at wavelengths comparable to that of the $S_0 \rightarrow S_1$ transition. A minimal model which can mimic this situation is an electronic three-level description (cf. Fig. \ref{fig:scheme}b). The first study of optical properties of linear three-level aggregates has been presented by \cite{knoester95:2780}. These authors showed that the coupling between local and nonlocal double excitations results in distinct features of the two-photon absorption spectrum. A region of coupling parameters has been identified which will lead to a strong distortion of the pump-probe spectrum. In fact under certain conditions (harmonic limit) the third-order response will vanish all together (\cite{knoester97:111}). This issue has subsequently been addressed for arbitrary aggregate geometries on the basis of the anharmonic exciton oscillator picture (\cite{kuhn96:8586}). In particular a time-domain four-wave mixing technique has been proposed to study the interplay between local and nonlocal double excitations (\cite{kuhn97:809}).

In the following we will specify the aggregate Hamiltonian, Eq. \re{eq:Hagg}, to the case of coupled three-level monomers. Restricting the summation indices to $a=(g,e,f)$ leads to a site-diagonal term which contains the non-adiabaticity operator mixing all states. However, in practice the non-adiabatic coupling to the ground state can be neglected, i.e. $\Theta_{m,ga}\approx 0$ ($a=(e,f))$. Concerning the coupling $\Theta_{m,fe}$ the restrictions of the three-level description becomes apparent (cf. Fig. \ref{fig:scheme}b). Normally, the state $\ket{\varphi_{mf}}$ will \emph{not} be the second excited state, but a certain higher excited state, S$_n$, which happens to have about the same transition frequency with respect to S$_1$ as the $S_0 \rightarrow S_1$ transition. In other words, there will be a sizable number of electronic states, which are energetically in between $\ket{\varphi_{me}}$ and $\ket{\varphi_{mf}}$ and it is rather likely that the electronic relaxation proceeds via these states. Since not much is known about these states and their couplings, $\Theta_{m,fe}$ has to be viewed as an \emph{effective} coupling.

Next we discuss the Coulomb integrals $J_{m n}(a b, c d)$, which can be classified as follows (\cite{may11}): First, there are interactions between charges ($J_{m n}(a a, a a)$, $J_{mn}(ab,ab)$) and between charges and transitions ($J_{m n}(a b, c a)$ or $J_{m n}(a b, b d)$). They contain the effect of aggregation on the adiabatic energy levels. If no first principles determination of the aggregate Hamiltonian is attempted, this effect is usually considered as being included in the definition of the adiabatic states. Note that these terms are also held responsible for the exciton-exciton interaction which can lead to the formation of bi-excitons (\cite{spano91:1400,juzeliunas:6916,knoester03:1}).
Second, there are matrix elements of the type $J_{mn}(aa,bb)$ that present interactions between simultaneous excitations and de-excitations of both monomers. From the energetic point of view this will lead to a rather off-resonant process. Neglecting these terms is usually called Heitler-London approximation. Within this approximation the number of excitations will be conserved. Finally, matrix elements of the type $J_{mn}(ab,ab)$  and $J_{mn}(ab,cc)$ which give the interaction between excitation and de-excitation of the monomers. In contrast to the previous case this interaction facilitates resonant EET and therefore it is often the only term that is considered in modeling of EET. For the electronic three-level system there are two types of matrix elements which need to be considered, i.e.
$J_{mn}(eg,eg)=J_{mn}^{(eg)}$ which facilitates resonant EET involving S$_0 \rightarrow $S$_1$ transitions and $J_{mn}(fg,ee)=J_{mn}^{(fe)}$ which involves simultaneous 
S$_0 \rightarrow $S$_1$ and S$_1 \rightarrow $S$_n$ transitions.

Using this approximation the aggregate Hamiltonian, Eq. \re{eq:Hagg}, can be recast into a simpler form. To this end 
a classification of the electronic states according to the number of excitations which are present, e.g., as a consequence of the interaction with the light field is introduced. For the description of third-order nonlinear optical spectroscopy it suffices to include the ground state
\bea
\label{c9-ground}
\ket{0} = \prod_{m} \ket{\varphi_{mg}} \; ,
\eea
the state carrying a single excitation
\bea
\label{c9-one-exc}
\ket{m} = \ket{\varphi_{me}} \prod\limits_{n \neq m} \ket{\varphi_{ng}} \; ,
\eea
and the one carrying two excitations which can be either local or nonlocal
\bea
\label{c9-two-exc}
\ket{m n} = (1-\delta_{mn}) \ket{\varphi_{me}} \ket{\varphi_{n e}}
\prod\limits_{k \neq m, n} \ket{\varphi_{kg}} + \delta_{mn} \ket{\varphi_{mf}}\prod\limits_{k \neq m} \ket{\varphi_{kg}} \; .
\eea
Introduction of this excitation expansion allows to rewrite the Hamiltonian as follows
\be
H= H^{(0)}+ H^{(1)} +H^{(2)} + H^{(1-2)} \, .
\ee
Here the ground state Hamiltonian contains just the contributions from nuclear motions
\be
\label{eq:h0L}
H^{(0)} = \sum\limits_m H_{m,g} \ket{0} \bra{0} \equiv \mathcal{E}_0 \ket{0} \bra{0} \, .
\ee
For the single excitation manifold we have
\be
H^{(1)} = \sum\limits_{m,n} [\delta_{mn} (\mathcal{E}_0  + U_{m,eg}) + J_{mn}^{(eg)}] \ket{m} \bra{n} \, ,
\ee
where 
\be
\label{eq:gap}
U_{m,ab} = V_{m,a} - V_{m,b}
\ee
is the nuclear coordinate dependent electronic gap coordinate, i.e. $U_{m,ab} (R_m)$
Finally, we have for the double excitation manifold
\bea
H^{(2)} &=& \sum\limits_{l \ge k}\sum\limits_{n \ge m}\{ \delta_{mk} \delta_{nl} [(1-\delta_{mn}) (\mathcal{E}_0  + U_{m,eg} +  U_{n,eg})
+\delta_{mn} (\mathcal{E}_0 + U_{m,fg} )] \nonumber\\
&+& (1-\delta_{mn})(1-\delta_{kl})[ J_{km}^{(eg)}  \delta_{nl}+J_{ln}^{(eg)}  \delta_{mk} +J_{kn}^{(eg)}  \delta_{lm} + J_{lm}^{(eg)}  \delta_{kn} ] \nonumber\\
& + &(1-\delta_{mn})\delta_{kl}[ J_{km}^{(fe)}  \delta_{kn}+J_{kn}^{(fe)}  \delta_{mk}]\nonumber\\
& + & \delta_{mn}(1-\delta_{kl})[J_{km}^{(fe)}  \delta_{lm}+J_{lm}^{(fe)}  \delta_{km}] \}\ket{kl} \bra{mn} \nonumber \, .
\eea
The non-adiabatic coupling has only on-site terms, i.e.
\be
\label{eq:HnaL}
H^{(1-2)} =\sum_m \Theta_{m,fe} \ket{mm} \bra{m} + {\rm h.c.}\, .
\ee
The same holds true for the electric dipole coupling to the laser field which reads
\bea
H_{\rm field}(t)& =& - \mathbf{E}(t) \sum_m \mathbf{d}_{m,eg} \ket{m}\bra{0} -\mathbf{E}(t) \sum_{n \ge m} \{ \delta_{mn} \mathbf{d}_{m,fe} \ket{mm}\bra{m} \nonumber\\
&+& (1-\delta_{mn})  [\mathbf{d}_{m,eg} \ket{mn}\bra{n}+ \mathbf{d}_{n,eg} \ket{mn}\bra{m}] \}+ {\rm h.c.} \,.
\label{eq:hfieldL}
\eea
Eqs. (\ref{eq:h0L}-\ref{eq:hfieldL}) provide a complete description of the laser-driven exciton dynamics up to doubly excited states.  By virtue of the electronic gap coordinates, Eq. (\ref{eq:gap}),  and the non-adiabatic coupling the Hamiltonian still depends on the nuclear degrees of freedom. This will lead to energy and phase relaxation processes within the excitation subspaces (intraband) as well as to transitions between them (interband). Leaving aside the non-adiabatic transitions and considering cases where the Coulomb coupling dominates the effect of the EVC, it is customary to introduce \emph{Frenkel exciton states} by separate diagonalization of the single $(N=1)$ and double $(N=2)$ excitation Hamiltonians for a fixed nuclear configuration (i.e. at the ground state equilibrium)
\be
H^{(N)} \ket{\alpha_N} = E_{\alpha_N}\ket{\alpha_N} \, .
\label{eq:He}
\ee
The exciton states can be decomposed into the basis of the local excitation states as follows:
\be
\ket{\alpha_1} = \sum_m C_{m,\alpha_1} \ket{m} \, ,
\ee
\be
\ket{\alpha_2} = \sum_{n\ge m} C_{mn, \alpha_2} \ket{mn} \, .
\ee
Notice that if $N_{\rm mol}$ is the number of monomers, one has $N_{\rm mol}$ one-exciton and $N_{\rm mol}(N_{\rm mol}+1)/2$ two-exciton states which are often called bands.

Using this expansion, the non-adiabatic coupling, Eq. \re{eq:HnaL}, and the aggregate-field interaction, Eq. \re{eq:hfieldL}, can expressed as couplings between different exciton manifolds.

In general, Eq. \re{eq:He}, has to be solved numerically. However, for special geometries analytical solutions are available. This includes the dimer, the linear open and closed chain (\cite{may11}) and certain tubular aggregates (\cite{knoester06:1}).  For example, for a homodimer with identical site energies $\hbar \omega_{eg}$ and coupling $J$, the eigenenergies of the one-exciton manifold are given by
\begin{equation}
\label{ }
E_{\alpha_1=\pm} = \frac{\hbar \omega_{eg}}{2} \pm J
\end{equation}
and the eigenstates are
\be
\ket{\alpha_1=\pm} = \frac{1}{\sqrt{2}}(\ket{1} \pm e^{-i{\rm arg}(J)}\ket{2} ) \,.
\ee
For the case of a linear aggregate of $N_{\rm mol}$ monomers having identical $S_0 \rightarrow$S$_1$ transition energies $\hbar \omega_{eg}$ and nearest-neighbor coupling $J$ one has
\begin{equation}
\label{ }
E_{\alpha_1} = \hbar \omega_{eg} + 2J \cos\left( \frac{\pi \alpha_1}{N_{\rm mol} +1}\right) \quad \quad \alpha_1=1,\ldots, N_{\rm mol}
\end{equation}
with the eigenstate expansion coefficients
\begin{equation}
\label{ }
C_{m,\alpha_1} = \sqrt{\frac{2}{N_{\rm mol} +1}} \sin\left( m \frac{\pi \alpha_1}{N_{\rm mol} +1}\right)
\end{equation}
Notice that linear aggregates are often considered by using periodic boundary conditions, which enables one to introduce the eigenstates as Bloch states having quasi-momentum  $k$  (\cite{knoester06:1}).
\subsection{Quantum Chemical Methods}
There are several strategies for the determination of the parameters of the Frenkel exciton Hamiltonian. On an empirical level, fitting to experimental data can be attempted. With increasing computer resources and the development of sophisticated methods for the solution of the electronic Schr\"odinger equation even for larger chromophores in solution, a first principles based  Frenkel exciton Hamiltonian can be constructed. Although semiempirical methods still enjoy popularity (\cite{fron08:1509}), in particular time-dependent density functional theory (TDDFT) plays a prominent role as it often gives a reasonable compromise between accuracy and numerical efficiency (for a critical overview see \cite{liu11:1971}). Here one can distinguish between two approaches: First, the aggregate is treated as a supermolecule, usually restricted to a representative dimer. Based on the splitting between excited states and their dependence on the dimer geometry, coupling parameters can be deduced. 

Second, the aggregate is viewed as being composed of monomers whose monomeric electron densities are interacting via the Coulomb potential. Thus one follows the line of the Frenkel Hamiltonian to combine monomer transition energies and dipoles with calculations of Coulomb matrix elements. There are several methods for obtaining $J_{mn}(ab,cd)$ in Eq. \re{eq:jmn}. Scholes and coworkers (\cite{krueger98:2284}) have been the first to propose a practical method for a first principles based calculation of the Coulomb coupling. Their TDC (transition density cube) approach coarse-grains the space for integration of the transition densities, thus expressing the interaction as a sum over pairs of TDCs. An alternative has been suggested by Renger and coworkers (\cite{madjet06:17268}). Their TrEsp (transition charge from electrostatic potential) method employs atomic partial charges in the summation of the Coulomb interaction. These partial charges are obtained by fitting the electrostatic potential associated with the monomeric transition densities. If $q^{(m)}_{I}(a,d)$ denotes the $I$th charge associated with the transition density $\rho_{ad}^{(m)}$ the Coulomb integral becomes
\begin{equation}
\label{ }
J_{mn}(ab,cd) \approx \sum_{I,J} \frac{q^{(m)}_{I}(a,d) q^{(n)}_{J}(b,c)}{|\mathbf{ R}_I -\mathbf{ R}_J|}
\end{equation}
where $\mathbf{ R}_I$ is the position of the partial charge. As compared to the TDC method this approach has the advantage that essentially converged results can be obtained using some tens of partial charges. More recently direct implementations of the calculation of the Coulomb coupling became available, which utilize the integration and pre-screening tools available in quantum chemistry packages (\cite{fuckel08:074505}).

Studying the higher excited states which are relevant for two photon transitions, one faces a serious computational challenge. The S$_1 \rightarrow$ S$_n$ transitions often involve transitions between molecular orbitals with double excitation character (\cite{pabst08:214101}) and therefore cannot be treated by TDDFT methods. Recently, we have shown that the MRCI/DFT (multi-reference configuration interaction) approach developed by \cite{grimme99:5645} can provide guidance for the assignment of S$_1 \rightarrow$ S$_n$ transitions in large chromophores such as perylene bisimides (\cite{ambrosek11:xxx}).
\subsection{Exciton-Vibrational Coupling}
\label{sec:evc}
The coupling between electronic and nuclear degrees of freedom leads to the appearance of the energy gap coordinate $U_{m,ab}$, Eq. \re{eq:gap}, in the monomer Hamiltonian. Furthermore, upon aggregation the Coulomb coupling between the monomers depends on the intermonomer distance and its variation, e.g., due to thermal fluctuations, will have an influence on the exciton dynamics. Besides theses intra-aggregate coordinates, there will be an influence on the degrees of freedom of the solvent environment. There are various strategies for incorporating the effect of EVC. Of considerable importance has been the  Haken-Strobl-Reineker model (\cite{haken72:253,haken73:135}) which incorporates EVC by means of adding stochastic modulations to the bare electronic transitions energies and Coulomb couplings. This implies  a certain model for the correlation functions of the energy gap coordinates, Eq. \re{eq:gap}, 
$\langle U_{m,eg}(t)U_{n,eg}(0) \rangle$, where the time-dependence is due to the dynamics of the nuclear degrees of freedom and the averaging is performed with respect to their equilibrium ensemble. More recently, these correlation functions became amenable to molecular dynamics simulations. Moreover, quantum-classical approaches have been developed to obtain the one-exciton Hamiltonian including the Coulomb coupling on-the-fly along classical molecular dynamics trajectories (\cite{zhu07:118,zhu08:154905,zhu08:117}).

A different perspective is obtained by introducing collective harmonic normal mode vibrations (\cite{renger01:137,may11}). For the purpose of illustration we consider the case where the dependence of the Coulomb coupling on the nuclear degrees of freedom can be neglected.
The introduction of normal modes enables an analysis in terms of oscillators whose equilibrium positions are shifted upon electronic excitation, a problem which to some extent facilitates an analytical treatment (Huang-Rhys model, see Fig. \ref{fig:scheme}). In case of a continuous distribution of oscillator frequencies, EVC is described by means of a spectral density. Such a model applies in particular to the low-frequency part of the oscillator spectrum. Without discriminating between localized (at some monomer) and delocalized modes, the EVC Hamiltonian is given by (\cite{may11})
\bea
\label{eq:hexvib}
H_{\rm ex-vib}&=& \sum_{m} \sum_\xi g_{m,e}^{(\xi)} q_\xi \ket{m} \bra{m} +\sum_{m,n}\sum_\xi [ \delta_{mn} g_{m,f}^{(\xi)} \nonumber\\
&+& (1-\delta_{mn})(g_{m,e}^{(\xi)}+g_{n,e}^{(\xi)})] q_\xi \ket{mn} \bra{mn} \, .
\eea
Here $\{ q_\xi\}$ denotes the set of vibrational normal modes and $g_{n,a}^{(\xi)}$ is the coupling constant for the $\xi$th mode and state $\ket{\varphi_{ne}}$. It follows from a linear expansion of the gap function, Eq. \re{eq:gap}, in terms of $q_\xi$.

The absorption spectra of the dyes, which are used for making molecular aggregates are typically characterized by a pronounced vibrational progression giving rise to a peak in the 1000-1500 \cm range to the blue of the 0-0 transition (see Fig. \ref{fig:PBImono}).  The underlying coupling to higher frequency intramonomer modes often needs to be accounted for on a different footing. Here a possible strategy is the separation of such intramolecular modes, $Q_{m,\xi}$, from the total set of aggregate normal modes which results in two EVC Hamiltonians plus some residual coupling between the (artificially) separated modes. The intramolecular modes can then be treated, for instance, non-perturbatively by defining respective vibronic exciton states. While this is straightforward for a dimer or trimer   (\cite{matro95:2568,kuhn96:99,seibt06:354,seibt07:164308,polyutov11:xxx}), the treatment of larger aggregates requires further approximations concerning possible excitations. For instance, building on earlier work by \cite{philpott71:2039}, Spano and coworkers employed a two-particle approximation to a complete multi-particle basis to develop a Frenkel exciton polaron theory (\cite{spano10:429}). Here a local vibronic excitation is "dressed" by vibrationally excited ground state monomers. An extension of the two-particle approximation has been proposed by Engel and coworkers using the multi-configuration time-dependent Hartree (MCTDH) ansatz for the correlated wave function (\cite{seibt09:13475}). However, due to numerical limitations this approach seems to be restricted to smaller aggregates. Still a different strategy has been followed by Briggs et al. who employed a Green's function approach in the so-called "coherent exciton scattering" (CES) approximation, which amounts to neglecting vibrational excitation in the electronic ground state (\cite{roden09:044909}).
\subsection{Dissipative EET Dynamics and Spectroscopy}
The EET dynamics in molecular aggregates is an example for a system with decoherence and dissipation (\cite{may11}). The coherent dynamics of the (vibronic) excitons is disturbed by the interaction with some environment (heat bath). Depending on the ratio between the Coulomb and system-bath couplings, the limits of (partly) coherent and incoherent transfer may be realized. The incoherent transfer limit under the assumption of strong intramonomer EVC is commonly known as F\"orster theory (\cite{foerster48:55}).
A unifying framework for the theoretical description is provided by density matrix theory  (\cite{renger01:137,may11}). Here a reduced density operator, $\rho(t)$, is defined by performing the trace of the total statistical operator with respect to the bath degrees of freedom. There are various methods for incorporating the system-bath interaction. Among the early approaches is the stochastic Haken-Strobl-Reineker model (\cite{haken72:253,haken73:135}) which assumes a Gaussian-Markovian behavior of the bath-induced fluctuations. While this is a high-temperature theory, a generalization to finite temperatures has been given by \v{C}apek (\cite{capek85:101}). More recently the non-Markovian case had been considered assuming a dichotomic behavior of the noise (\cite{chen11:5499}).
The assumption of a stochastic influence of the environment is also present in the stochastic Schr\"odinger equation approach (\cite{roden09:058301}). While in most applications model assumptions concerning the  fluctuations of transition energies and Coulomb couplings are introduced, they can be also obtained from quantum-classical simulations  (\cite{zhu07:118,zhu08:117,zhu08:154905}).

In the weak system-bath coupling limit multilevel Redfield theory formulated in terms of exciton eigenstates can be used, which is based on second-order perturbation theory assuming a Markovian dynamics (\cite{renger01:137,may11}). Here, the bath enters via its spectral density which can be obtained, e.g., from fitting experimental spectra (\cite{kuhn02:15}). The approximations of Redfield theory are lifted within the path integral (\cite{thorwart09:234}) and the hierarchy equation (\cite{ishizaki05:3131}) approaches, what comes at the expense of a restriction to small aggregates.

To connect the quantum dynamics with spectroscopy, the macroscopic polarization needs to be calculated, which for a homogenous sample with aggregate density $n_{\rm agg}$ is given by $\mathbf{ P}(t)=n_{\rm agg} {\rm tr}(\mathbf{ d} \rho(t))$, where the trace is with respect to the relevant system degrees of freedom and $\mathbf{ d}$ is their dipole moment operator. In the weak field case one can expand the reduced density operator $\rho(t)$ in powers of the external field to obtain the optical response in terms of linear and nonlinear response functions, i.e. (multi-)time correlation functions of the dipole moment operator (\cite{mukamel95}). The response can also be expressed in terms of Green's functions, e.g. (\cite{chernyak95:213,mukamel04:2073,roden08:258}) or sum-over-states expressions (\cite{knoester97:111}). A related approach is the so-called nonlinear exciton equation of motion method (\cite{axt98:145,mukamel04:2073}). Here a set of Heisenberg equations of motion for exciton creation and annihilation operators is derived and numerically solved in the time-domain. The advantage is that one obtains a clear picture concerning the elementary excitation contributing to a particular nonlinear signal. Further, being formulated in the local monomer basis, it scales favorably with aggregate size. The drawback is in the approximate description of dissipation and coherence and in the necessity to find appropriate truncation schemes for the many body hierarchy of equations of motion (\cite{axt98:145,mukamel04:2073}). While originally being formulated for electronic two-level monomers only, higher excited states (e.g. three-level model) can be incorporated within the frame of the deformed boson theory (\cite{kuhn96:8586}). 
Increasing the field strength, the perturbative response function approach becomes cumbersome. In order to keep track of the field order and signal propagation direction, a carrier wave expansion of the density operator can be performed what leads to a hierarchy of equations for some auxiliary density operators (\cite{renger01:137}).

\subsection{Exciton-Exciton Annihilation} 
\label{sec:EEA}
It has been known for decades that annihilation of excitons in molecular aggregates is an important loss mechanism in EET. In luminescence experiments it leads to a reduction of the quantum yield and lifetime, thus restricting applications in optical devices. Early work by \cite{suna70_1716} on molecular crystals focussed on the derivation of the bimolecular annihilation rate $\gamma$ in the equation for the exciton density $n(\mathbf{ r},t)$ given by (cf. related work by \cite{kenkre80_2089})
\begin{equation}
\label{eq:anni}
\frac{\partial n(\mathbf{ r},t)}{\partial t} = -\gamma n^2(\mathbf{ r},t) \,. 
\end{equation}
Subsequent work on linear aggregates by Malyshev and coworkers emphasized the role of higher excited electronic monomer states and golden rule type annihilation rates have been derived (\cite{malyshev99_117,malyshev00:31}). This problem has also been cast into a nonlinear exciton equation of motion approach based on a local site representation (\cite{renger00:807}). A consequent multi-exciton density matrix formulation of EEA has been given by May and coworkers (\cite{renger01:137,bruggemann03:746}). Invoking some approximations the two-exciton ($\alpha_2$) to one-exciton ($\alpha_1$) interband relaxation rate can be cast into the simple form (\cite{bruggemann03:746})
\begin{equation}
\label{eq:EEA}
k^{\rm (EEA)}_{\alpha_2 \rightarrow \alpha_1} = \sum_m |C_{m,\alpha_1} C_{mm,\alpha_2}|^2 k^{\rm (IC)}_m \, .
\end{equation}
Here, $k^{\rm (IC)}_m$ is the monomeric internal conversion rate for site $m$. Although analytic expressions for this rate can be given in terms of matrix elements of the non-adiabatic coupling operator, Eq. \re{eq:HnaL}, any first principles calculation is hardly possible due to the fact that very highly excited electronic states will be involved. An important point of Eq. \re{eq:EEA} is that the annihilation rate is proportional to the overlap of the local exciton wave function coefficients in the aggregate. This does not come unexpectedly in view of the local character of the internal conversion process. Recently, EEA has been studied from the perspective of higher order perturbation theory of EET (\cite{may09:10086}).

%
\begin{figure}
\begin{center}
\includegraphics[width=0.7\textwidth]{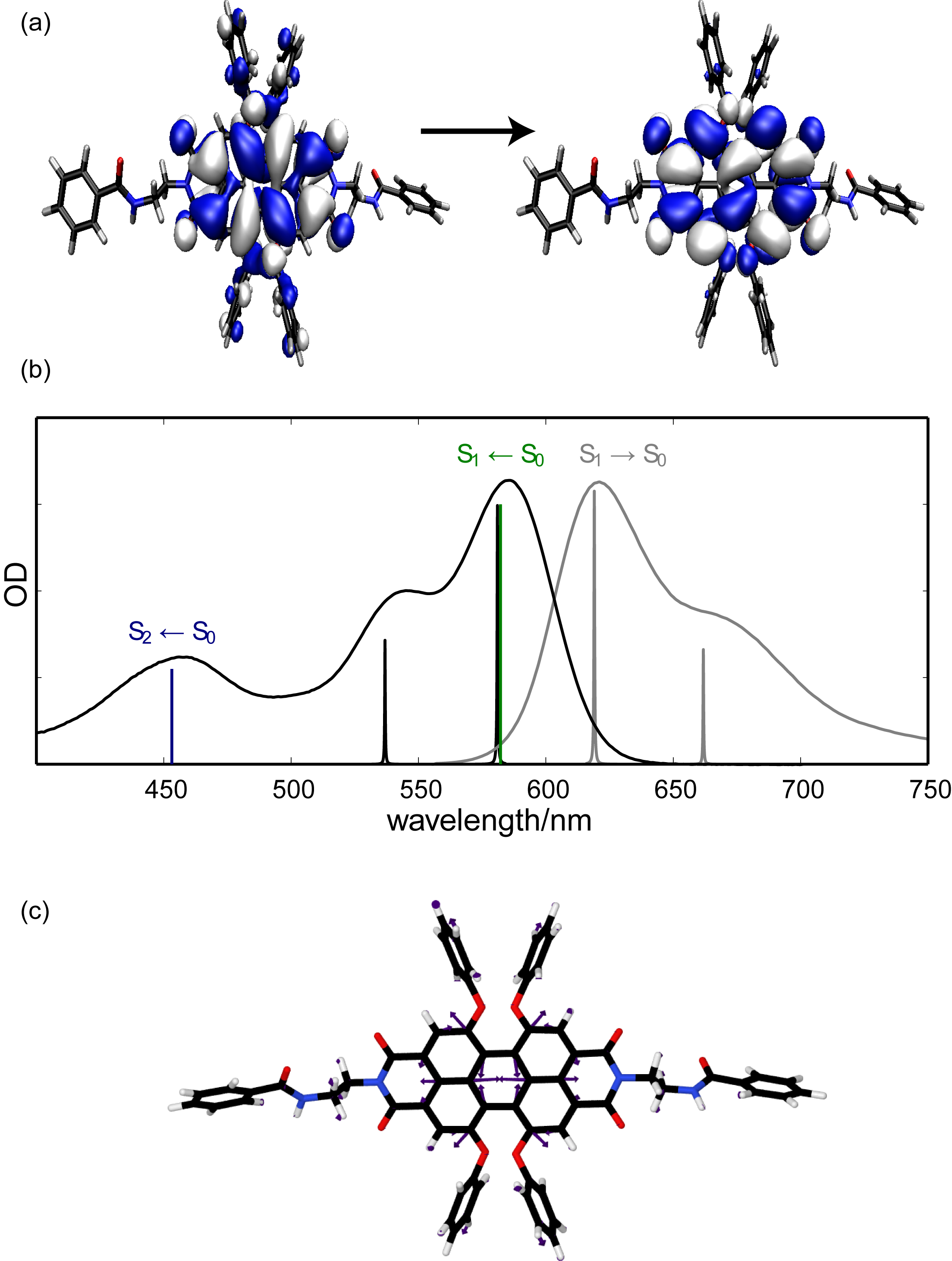}
\caption{PBI-1 ($N$,$N$'-Di($N$-(2-aminoethyl)-benzamide)-1,6,7,12-tetra(4-tert-butyl\-phe\-noxy)-
3,4:9,10-perylene\-biscarboximide) monomer absorption spectrum and calculations for a reduced system: (a) leading HOMO-LUMO transition contributing to the S$_0 \rightarrow $ S$_1$ absorption, (b) absorption and fluorescence spectrum and calculated results (narrow lines), (c) vibrational mode that leads to the vibrational progression for the electronic S$_0 \rightarrow $ S$_1$ absorption (\cite{ambrosek11:xxx})}
\label{fig:PBImono}
\end{center}
\end{figure}
%

%
\section{Applications}
\label{sec:appl}
\subsection{Aggregation of Dye Molecules}

The first observations of aggregates were made by \cite{Schei36,Schei37} and \cite{Jel36,Jel37} on pseudoisocyanine dyes. Since then aggregation has been found to occur in a large number of molecular systems. Several excellent review articles provide overviews about the relevant compounds and their specific aggregation behavior (\cite{mobius95:437,Boh93,wuerthner11:3376}). Cynanine dyes are probably the most popular group of molecules to study aggregates. Various cationic carbocyanine (\cite{Sche01,Bur95,Nem09}), oxa- and thiacyanine (\cite{Kuh71}), pseudoisocyanine (PIC) (\cite{sundstrom88:2754,Fid91}), and neutral merocyanine dyes (\cite{Ham02}) have been investigated. Chlorophyll dyes and structurally related compounds have been intensively studied due to their close relationship with natural light harvesting systems (\cite{You05,Bal05,Mis99}). During the last decade aggregates made from perylene bisimides (PBI) have attracted increasing interest, since these dyes exhibit high quantum yields and good photostability and are versatile building blocks for functional supramolecular structures (\cite{Kai07,Li08,Mar11}). 

In Fig. \ref{fig:PBImono} we show the analysis of the absorption spectrum of the perylene bisimide dye molecule $N$,$N$'-Di($N$-(2-aminoethyl)-benzamide)-1,6,7,12-tetra(4-tert-butyl\-phe\-noxy)-3,4:9,10-perylene\-biscar\-bo\-xi\-mide (PBI-1) (\cite{ambrosek11:xxx}), which includes the S$_0 \rightarrow$ S$_1$ as well as the S$_0 \rightarrow$ S$_2$ transitions. The transitions can be characterized as being of $\pi\pi^*$ type with the $\pi$ electron density being located at the central PBI core with some contribution on the bay substituents. An important feature of the absorption spectrum is the shoulder on the blue side of the S$_0 \rightarrow$ S$_1$ transition which has been found to be due to a Franck-Condon progression. A normal mode vibration which contributes to this progression is shown in panel (c) of Fig. \ref{fig:PBImono}. Such signatures of electron-vibrational coupling are typical for many chromophores used for building molecular aggregates. It has strong implications also for the aggregate spectra and dynamics as will be shown below.
\begin{figure}
\begin{center}
\includegraphics[width=0.75\textwidth]{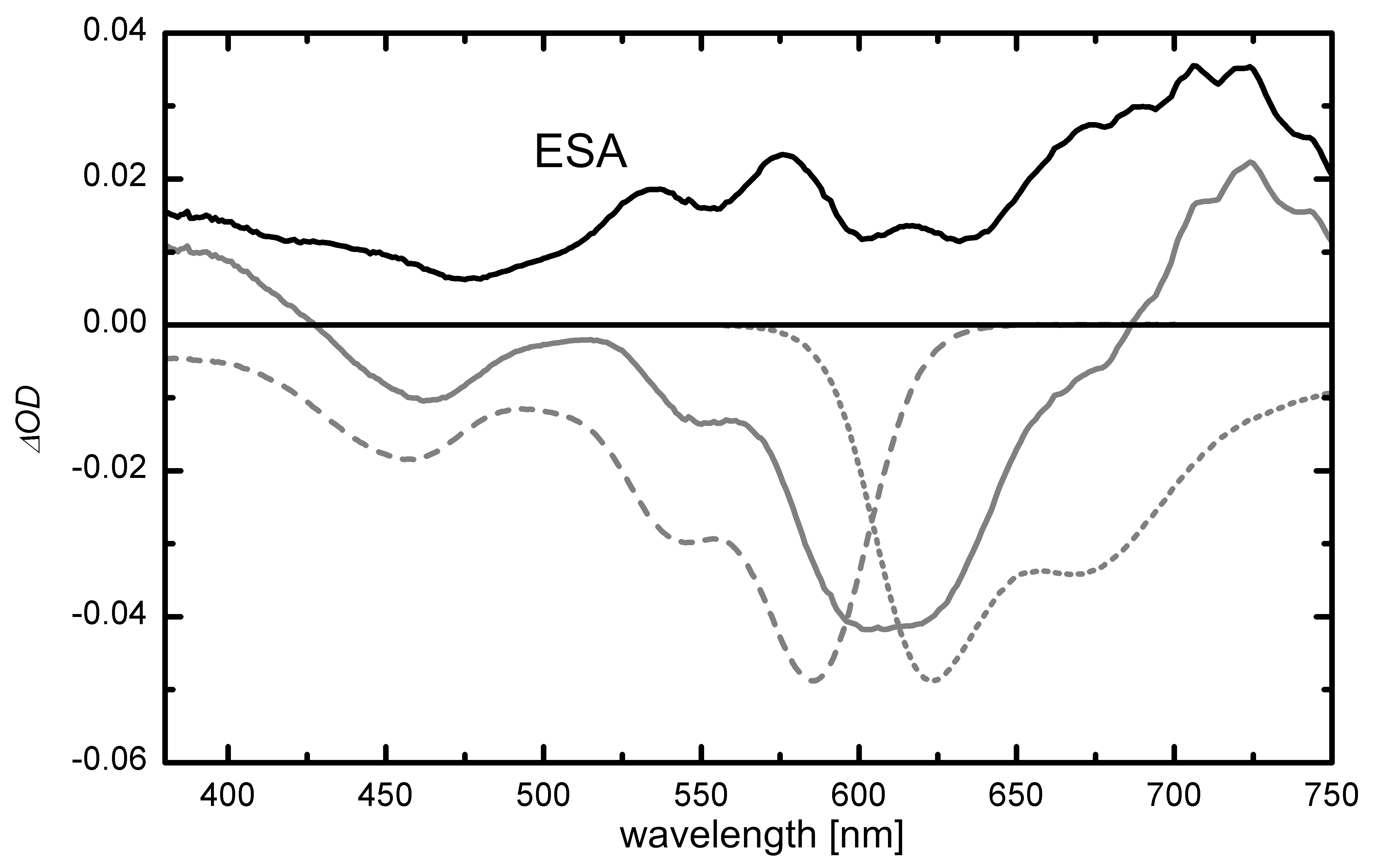}
\caption{PBI-1 monomer excited state absorption (ESA) spectrum, S$_1 \rightarrow $ S$_n$, which has been obtained by subtracting the ground state bleach (dashed grey) and stimulated emission (dotted grey) bands from the transient absorption (solid grey) spectrum measured at a delay time of 100 ps in dichloromethane  (for more details, see \cite{ambrosek11:xxx}).}
\label{fig:esa}
\end{center}
\end{figure}

In the context of nonlinear spectroscopy but also for the understanding of EEA the characterization of higher electronic states, S$_n$, is important and in particular in the range of about twice the S$_0 \rightarrow$ S$_1$ transition energy. This spectral range is accessible by employing transient pump-probe spectroscopy (cf. Section \ref{sec:ts}) from which ESA profiles can be extracted. However, the actual procedure requires detailed knowledge of the ground state bleach and stimulated emission spectra for subtraction from the total signal. If this cannot be obtained spurious features will appear in the excited state absorption which facilitates a semi-quantitative discussion of the spectrum only. As an example we show the ESA of PBI-1 in Fig. \ref{fig:esa}, where at most two broad bands can be identified that extend from 500 and 600 nm and from 650 to 750 nm. The theoretical assignment of these spectral features is hampered by the lack of accurate methods for highly excited states of these large organic molecules. For instance, using the DFT/MRCI approach, but for  a reduced model of the PBI-1 monomer (axial groups replaced by hydrogens), transitions in this range are found at 861 nm, 665 nm, and 594 nm (\cite{ambrosek11:xxx}). 

\begin{figure}
\begin{center}
\includegraphics[width=0.6\textwidth]{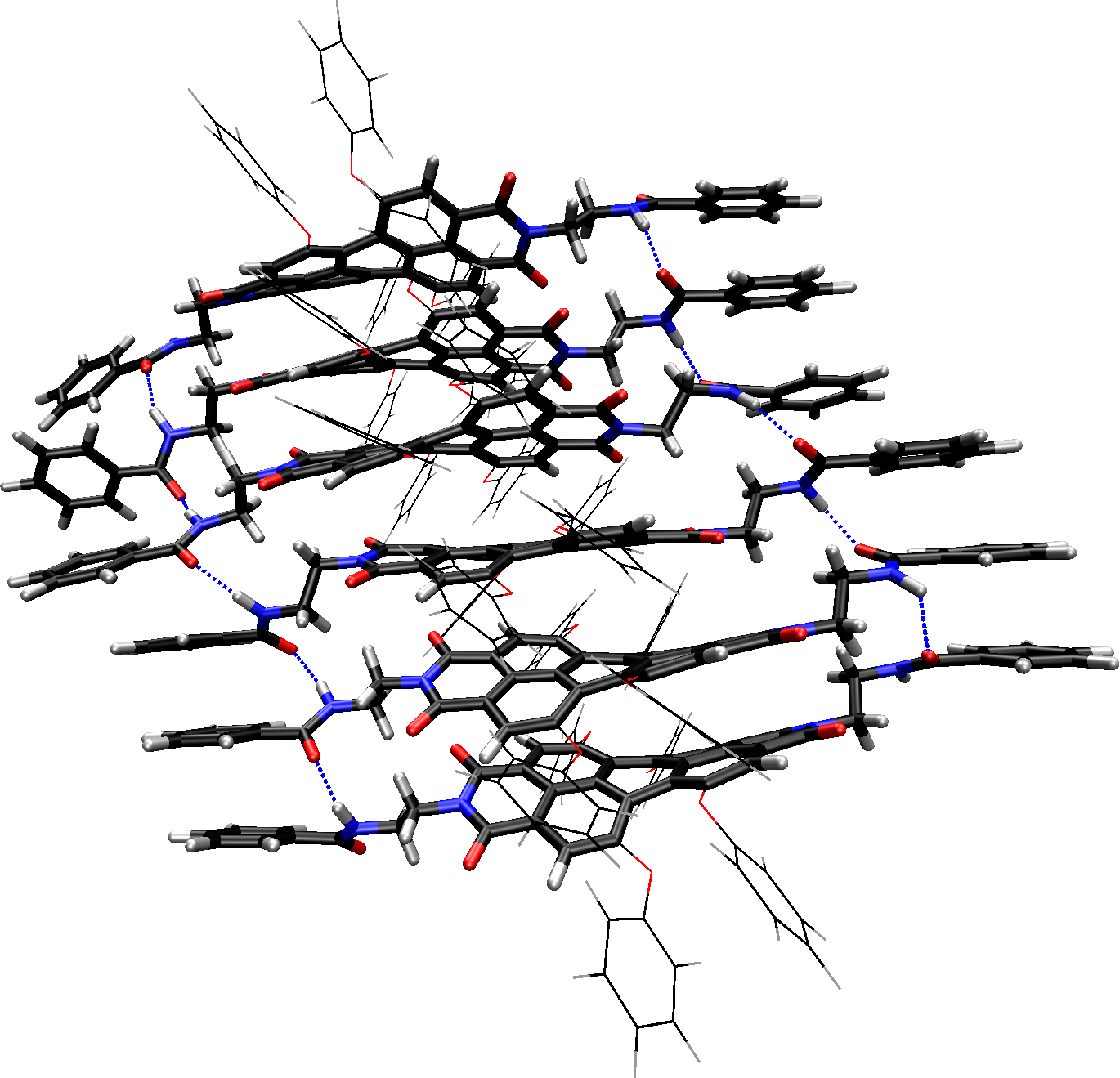}
\caption{Structure of a PBI hexamer obtained from geometry optimization using the tight-binding DFT method. Clearly visible are the hydrogen bonds, which give some stabilization of the structure in addition to the van der Waals interaction of the $\pi$ systems of the chromophores  (\cite{ambrosek11:yyy}).}
\label{fig:hex}
\end{center}
\end{figure}
The attraction between the monomers of dye aggregates results from van-der-Waals interaction between the extended $\pi$-systems of the chromophores and from intermolecular hydrogen bonds. In particular the later are used to control the structure of the aggregates which is on the other side strongly influenced by steric effects. To obtain a detailed notion of the aggregate structure is a nontrivial task, but to some extent can be guided by Computational Chemistry. Exemplarily we show the result of a structural optimization of a PBI-1 hexamer in Fig. \ref{fig:hex}.

Many cyanine dyes seem to form a brick stone structure, at least when monolayers are prepared (\cite{Dus88,mobius95:437}). For PIC aggregates investigations over several decades indicate that double stranded mesoaggregates are formed over which the excitons can be completely delocalized, at least at low temperatures (\cite{Dal74}). The mesoaggregates can build up fibers which arrange in complex bundles at high concentrations (\cite{Ber00,Ber02}). However, the exact structure depends on temperature, solvent properties, counter ions and probably also on the preparation procedure.

\subsection{Steady State Absorption and Fluorescence of Aggregates}
\begin{figure}
\begin{center}
\includegraphics[width=0.65\textwidth]{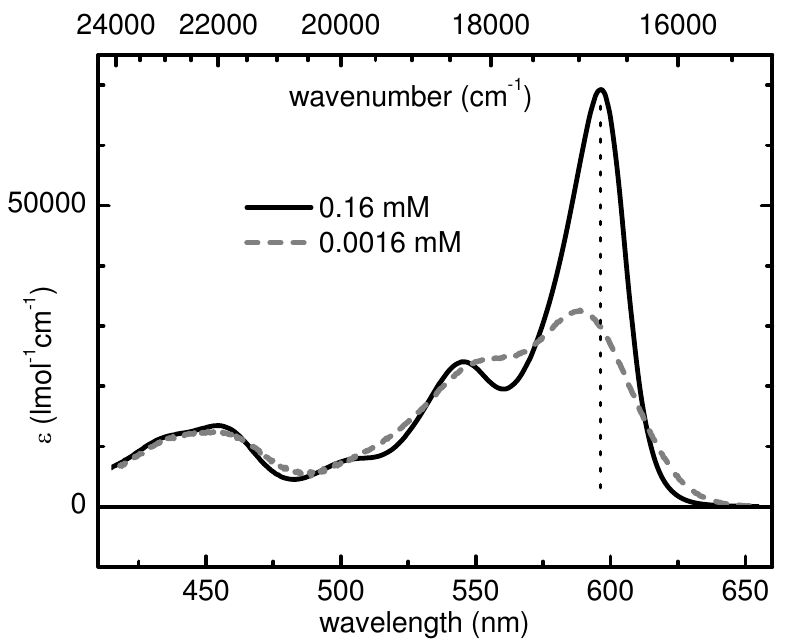}
\caption{Spectral changes due to the formation of J-aggregates of a core-substituted perylene bisimide dye (\cite{Li08,Mar11}).}
\label{fig:abs}
\end{center}
\end{figure}

Aggregation is in most cases demonstrated by changes in the stationary absorption spectrum in comparison to the monomer in highly diluted solutions. In the case of formation of J-aggregates a sharp band, red shifted with respect to the first electronic absorption band of the monomer is observed which rises with increasing dye concentration. In J-aggregates the slip angle between the monomeric transition dipole moments and the intermolecular connection axis is below the magic angle of 54.7$^\circ$ (cf. Eq. \re{c9-v12dd}) and the lowest exciton state carries most of the oscillator strength (\cite{Boh93,Bur95}). Figure \ref{fig:abs} shows the changes in the case of a PBI aggregate (\cite{Li08,Mar11}). The lowest band gains additional absorption strength, experiences a slight red shift, and becomes narrower. Here the delocalization length is only about two monomers (see below). For large coherence domains very sharp absorption bands can result (\cite{mobius95:437}).

H-aggregates exhibit broad absorption bands which are blue shifted with respect to the monomer absorption since here the slip angle is larger than the magic angle and the upper excitonic states are responsible for the absorption (\cite{Boh93,wuerthner11:3376}).

\begin{figure}
\begin{center}
\includegraphics[width=0.5\textwidth]{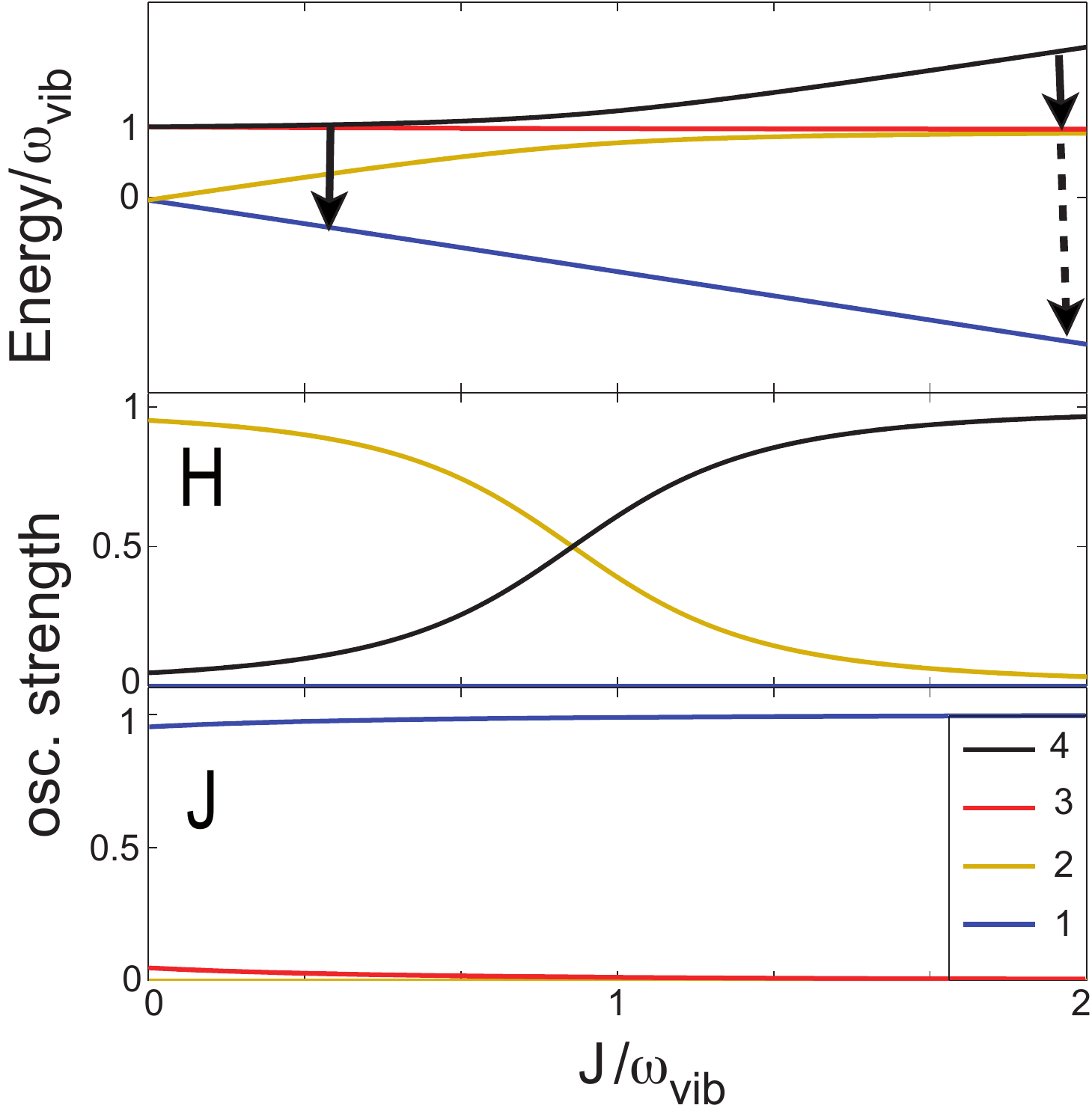}
\caption{Upper panel: Energy level structure of the one-exciton states, $\alpha_1=1,\ldots,4$, for a vibronic dimer as a function of the Coulomb coupling. The arrows indicate possible intraband transitions as discussed in Fig. \ref{fig:VD_pop}. Lower panels: Oscillator strengths for one-exciton transitions in H- and J-aggregate configuration (\cite{polyutov11:xxx}).}
\label{fig:VD_lev}
\end{center}
\end{figure}

In Fig. \ref{fig:VD_lev} we have plotted results on the energy level structure and oscillator strengths distribution for a vibronic dimer as a function of the Coulomb coupling strength (\cite{polyutov11:xxx}). The model includes the possibility of excitation of one vibrational quantum in the electronic excited state of each monomer. For small coupling strengths ($J/\omega_{\rm vib}$) J/H-aggregates have most oscillator strength in the transition to the lowest/highest state corresponding to the 0-0 transition. For J-aggregates this holds true also for increasing $J/\omega_{\rm vib}$, whereas for H-aggregates the exciton-vibrational state which carries oscillator strength is strongly mixed with respect to electronic and vibrational contributions. This simple model already hints at the more complex absorption line shape for H-aggregates as compared to J-aggregates (\cite{eisfeld06:376,spano10:429}).

In several cases a splitting of the J-band in two or more components is observed and attributed to the coexistence of two or more locally slightly different arrangements (\cite{Coo07,Fid91}). If the structure of aggregates consists of unit cells containing two molecules on non-equivalent sites Davydov splitting occurs. In the case of 3,3'-bis(sulfopropyl)-5,5'-dichloro-9-ethylthiacarbocyanine (THIATS) this results in the simultaneous appearance of J- and H-bands in the absorption. Excitation of the higher lying H-band leads to J-type fluorescence demonstrating that interband relaxation from the H- to the J-states takes place (\cite{Sche00}).

The linewidth of the J-band is strongly reduced compared to the monomer absorption (see above). This results from motional narrowing caused by the delocalization of the excitons over many monomers (\cite{knapp84:481}). The width is roughly $N^{-1/2}$ times the monomer width where $N$ is the number of monomers within the coherence domain. Applying this relation to PIC aggregates a coherence domain of about 60 monomers was estimated (\cite{knapp84:481}). The domain size is  often also called coherence length since the electronically excited states of chromophores within such a domain are coherently coupled. Numerical simulations of the absorption band taking energetic disorder into account indicate an exciton delocalization over about 100 molecules (\cite{Fid91}).

Another elegant way to characterize the number of coherently coupled chromophores is saturation spectroscopy with short laser pulses. From the saturation fluence $I_{\rm sat}=h \nu/ \sigma$ one can determine the absorption cross section $\sigma$ of the absorbing entity and can compare it to the cross section of the monomer. The short pulse duration ensures that the ground state is not repopulated during the excitation. In this way the coherence domain was estimated to be about 100 monomers for 1,1'-diethyl-2,2'-cyanine chloride (PIC-Cl) in water (\cite{Kop82}) and about 340 for 1,1'-diethyl-2,2'-cyanine bromide (PIC-Br) in a mixture of water and ethyleneglycol (\cite{Kob94}).

The fluorescence originates from the lowest exciton state since intraband relaxation is very efficient (see below). Accordingly, the fluorescence spectrum is a mirror image of the J-absorption band (\cite{mobius95:437}). In an ideal J-aggregate the Stokes shift should be zero since the lowest exciton state carries most of the oscillator strength and motional narrowing suppresses the impact of optical active vibrations. Energetic disorder leads in fact to a distribution of optical accessible excitonic states. However, the width of this distribution is typically narrow and in many cases extremely small Stokes shifts are observed like e.g. in PIC aggregates (\cite{Fid91}). In the case of weakly fluorescent dye molecules like PIC-Cl the formation of J-aggregates can drastically increase the fluorescence yield (\cite{Kat40}). This results from the increase of the transition dipole due to coupling of many chromophores and probably also from the reduced flexibility of the monomer within the aggregate which can suppress internal conversion pathways.
The excitons are mobile and can change between different delocalization domains. This results in an overall relaxation towards the lowest lying exciton states and can cause a significant Stokes shift of the fluorescence (\cite{Sche01}). The mobility of the excitons can be nicely demonstrated by incorporating small amounts of molecules in the aggregate which function as energy acceptors for the excitons. In this case extremely efficient quenching of the aggregate fluorescence occurs. It was found in PIC aggregates that one quencher molecule can deactivate 10$^3$ to 10$^6$ chromophores (\cite{Schei39}) exceeding by far the size of a coherence domain. This shows that an exciton can visit several domains within its lifetime. In bleaching experiments applying optical near-field microscopy an upper limit for the exciton diffusion length of 50~nm was found at room temperature for aggregates made from the dye 1,1'-diethyl-2,2'-cyanine iodide (PIC-I) (\cite{Hig95}).
Incorporation of electron accepting or donating molecules in an aggregate can also efficiently quench the aggregate fluorescence (\cite{Pen82}). If both kinds of molecules are incorporated electron transfer from the donor to the acceptor can occur via the aggregate. In this way it was demonstrated that charges can migrate up to 4~nm in aggregated monolayers of PIC dyes.

The fluorescence of H-aggregates is usually weak or totally lacking (\cite{Her77}). This results from the fact that the optically accessible exciton states are in the upper region of the exciton band and their population is quickly transferred to low-lying exciton states which possess no transition dipole to the ground state (\cite{Fink08}). Interestingly, excitons in nanowires made from H-aggregates of perylene tetracaboxylic diimide seem to exhibit a reasonable mobility. In these wires defects imprinted by a scanning microscope tip lead to effective quenching which points to an exciton diffusion length of at least 250~nm (\cite{Cha11}).

\subsection{Exciton Lifetime}

The lifetime of excitons in aggregates depends on various factors. As mentioned above the formation of J-aggregates can suppress nonradiative decay channels of the monomer. However, the radiative lifetime of J-aggregates is strongly reduced compared to the monomer since the coherent coupling of many chromophores results in a large transition dipole moment and the radiative rate is proportional to the number of chromophores in a coherence domain (\cite{Boh93,Bur95}). This effect is often called superradiance (\cite{knoester03:1}). The increase of the radiative rate due to aggregation was used to estimate the coherence length of zinc chlorin aggregates which are models for light harvesting systems to 10-15 chromophores (\cite{Pro02}).

The radiative rates of H-aggregates are very low since the transition of the lowest exciton state to the ground state is dipole forbidden (see above). The exciton lifetimes are usually limited by nonradiative processes. H-aggregates of perylene tetracaboxylic diimide show a lifetime increase from 1.7~ns to 46~ns upon cooling from room temperature to 25~K (\cite{Cha11}). This indicates that the exciton decay is caused by a thermally activated nonradiative process.

\subsection{Temperature Dependence}

The temperature influences the excitonic properties mostly in two ways. Firstly, with increasing temperature geometric and energetic fluctuations and phonon scattering decrease the size of coherence domains and reduce the number of effectively coupled monomers. Secondly, if the temperature is very low hops between different domains are suppressed. Then, in case of significant static disorder, the excitons can  not reach the lowest excitonic states but are trapped at local minima (\cite{Sche01}). 

In the case of Langmuir-Blodgett films made from 1-methyl-1'-octa-decyl-2,2'-cyanine perchlorate the fluorescence lifetime $\tau$ changes from 8.2~ps at room temperature to 5.5~ps at 143~K (\cite{Dor87}). The data were interpreted applying the equation $1/\tau = k_{\rm rad}(1+\beta T)$ (\cite{Moe79}). Here, $k_{\rm rad}$ is the radiative rate of the aggregate and $T$ the temperature. $\beta$ is a characteristic parameter of the particular dye, which parameterizes the nonradiative decay rate $k_{\rm nr}=k_{\rm rad} \beta T$  and describes its reduction with decreasing temperature. The data evaluation indicates that the size of the coherence domains increases by a factor of about 3 upon cooling from room temperature to 143~K (\cite{Dor87}). 

For J-aggregates of THIATS a minimum of the Stokes shift at intermediate temperatures of about 110~K and a complex temperature dependence of the fluorescence bandwidth were found (\cite{Sche01}). The relative Stokes shift which is the Stokes shift divided by the bandwidth of the fluorescence shows an interesting behavior. It is constant at very low temperatures up to 30~K, decays then exponentially with temperature, and adopts at about 80~K a pretty temperature independent value again. This was interpreted as an indication that at low temperatures the excitons are trapped at local energetic minima (\cite{Sche01}). At intermediate temperatures the hopping probability and the mobility of the excitons increases more and more with temperature till global minima can be reached. A similar behavior was also observed in core-substituted PBI aggregates (\cite{Kai09}). There the critical temperatures were  a factor of 2.8 higher than for THIATS pointing to a higher degree of disorder. The nonmonotonic temperature behavior of the Stokes shift due to the interplay between local bandstructure determined by the static disorder and relaxation dynamics has also been investigated theoretically  by  \cite{bednarz03:217401}.

\subsection{Two-Photon Spectroscopy of the Two-Exciton Band}
\begin{figure}
\begin{center}
\includegraphics[width=1.0\textwidth]{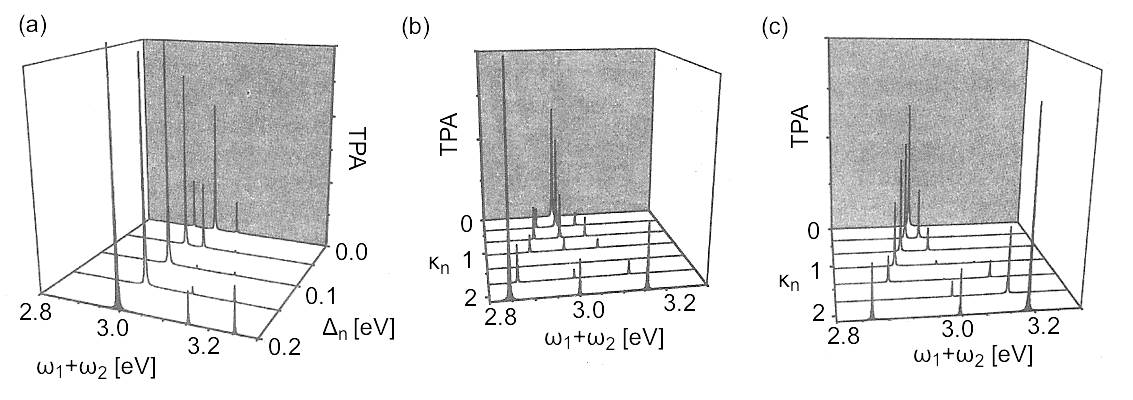}
\caption{Dependence of the two-photon absorption (TPA) spectrum of a heterodimer (transition energy difference is 0.05 eV, $J=-0.05$ eV) composed of electronic three-level systems on the local anharmonicity ($\Delta_n$) and the ratio of transition dipole moments ($\kappa_n$). (a) $\kappa_n=0.5$, (b) $\Delta_n=0$, (c) $\Delta_n=0.05$ eV (after \cite{kuhn96:8586}).}
\label{fig:tpa}
\end{center}
\end{figure}

The two-exciton band is characterized by an interplay between local and nonlocal double excitations. Their mixing depends on two parameters, i.e. the local anharmonicity of the electronic transitions, $\Delta_n = \hbar \omega_{n,ef}-\hbar \omega_{n,eg}$, and the ratio between transition dipole moments between S$_0 \rightarrow$ S$_1$ and S$_1 \rightarrow$ S$_n$ transitions, $\kappa_n$ (assuming that they have the same orientation). In Fig. \ref{fig:tpa} we show calculated two-photon absorption spectra of the two-exciton band for a heterodimer composed of three-level monomers.  
Increasing the anharmonicity parameter (panel (a)) the eigenstates forming the two-exciton band become less mixed and one can identify the delocalized double excitation around $\omega_1+\omega_2=3$ eV and the two local double excitation at higher energies. Increasing the ratio $\kappa_n$ leads to an increased mixing between the eigenstates causing level splitting since $J^{(ef)}_{mn} \propto d_{m,fe}d_{n,eg} = \kappa_m d_{m,eg}d_{n,eg}$.
Finally we notice that the nonlinear response vanishes if the system is harmonic, i.e. when $\Delta_n=0$ and $\kappa_n=\sqrt{2}$ (panel (b)). 
\subsection{Transient Spectra}
\label{sec:ts}
\begin{figure}
\begin{center}
\includegraphics[width=0.5\textwidth]{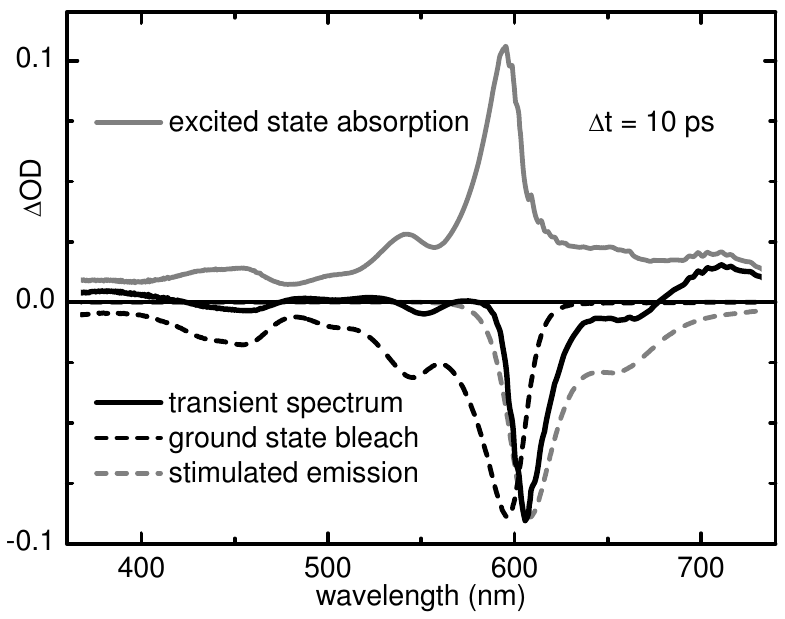}
\caption{Transient absorption spectrum (black line) 10~ps after optical excitation of PBI aggregates and its decomposition in ground state bleach, stimulated emission, and excited state absorption.}
\label{fig:TA}
\end{center}
\end{figure}

Ultrafast pump-probe spectroscopy is very helpful to characterize excitonic properties of aggregates. Figure \ref{fig:TA} shows a transient absorption spectrum obtained 10~ps after optical excitation of core-substituted PBI aggregates (\cite{Mar11}). The strong negative feature is due to the bleach of the ground state absorption and stimulated emission from the electronically excited state (see broken lines in Fig.~\ref{fig:TA}). Its sharp blue wing results from a partially hidden photoinduced absorption. It can be extracted by subtracting the bleach and the stimulated emission from the transient spectrum. The obtained ESA (grey line) shows a dominant band which is in shape and intensity similar to the low energy band of the ground state absorption but slightly blue shifted. This is the signature of a one- to two-exciton transition. The corresponding ESA band in aggregates of PIC-Br was directly observed at 1.5~K due to the very narrow bandwidth (\cite{Fid93}). The weaker bands at shorter wavelengths in Fig.Ê\ref{fig:TA} result from the vibronic structure of this transition while the absorption band at longer wavelengths is due to an ESA that is also observed in transient absorption measurements of the monomer (cf. Fig.~\ref{fig:esa}).

The difference, $\Delta E$, between the transition energies of the ground state absorption and the transient absorption band caused by the one- to two- exciton transition can be approximated for linear aggregates by  (\cite{Bur95})
\begin{equation}
\label{ }
\Delta E \approx \frac{3\pi^2 J}{{(N+1)^2}}
\end{equation}
Corresponding simulations of the transient absorption spectrum for aggregates made from 5,5',6,6'-tetrachloro-1,1'-diethyl-3,3'-di(4-sulfobutyl)benz\-imidazolo\-carbo\-cya\-nine  (TDBC) reveal at room temperature a delocalization length of about $N=15$ monomers (\cite{Bur95}). At a temperature of 1.7~K a delocalization length of 50-100 was estimated from transient absorption data for PIC-Br aggregates (\cite{Dur94}). For a merocyanine dye which forms two dimensional aggregates the coherence domain at 4.2~K was characterized by a combined analysis of the stationary and the transient absorption spectrum (\cite{Ham02}). The results indicate that the domain extends over 26 molecules in one direction and over 40 in the other direction.

Finally, we point out that effects of coherent exciton motion and dephasing  can also be detected as distinct features in ultrafast pump-probe experiments (\cite{heijs07:230}).

\subsection{Photon Echo Experiments}
Photon echo experiments are suitable to investigate line broadening mechanisms acting on the J-band. In the temperature range from 1.5~K to 100~K the photon echo decay measured at PIC-Br aggregates accelerates from 14~ps to about 4~ps (\cite{DEB87}). This was attributed to dephasing caused by the coupling to a low-frequency mode. However, even at 1.5~K the photon echo decay is faster than the fluorescence lifetime (\cite{Fid91}). It was argued that several excitonic states contribute and interference effects reduce the photon echo signal in time.
The decay of the two-pulse photon echo signal of TDBC aggregates at room temperature was analyzed simultaneously with the stationary J-absorption band using a fluctuation correlation function with two components described by the magnitudes ($\Delta$) and time scales ($\Lambda^{-1}$)  (\cite{Bur95})
\begin{equation}
\langle U_{n,eg}(t)U_{n,eg}(0) \rangle = \Delta^2_{\rm slow} e^{-\Lambda_{\rm slow} t} + \Delta^2_{\rm fast} e^{-\Lambda_{\rm fast} t} \, .
\end{equation}

The fast component, $\Lambda_{\rm fast}$, was found to be in the range of 5~fs with an amplitude of 10~THz and the slower one, $\Lambda_{\rm slow}$, in the range of 10~ps with an amplitude of 54~THz. Overall, the dephasing is slower than for dye monomers in solution indicating that the delocalization of the excitons over several chromophores averages out the fluctuations.

\subsection{Intraband Relaxation}
Each excitonic band is a dense manifold of excitonic states characterized in case of linear aggregates by their quasi-momentum $k$. Higher lying $k$-states can be populated by photo-excitation. Relaxation to lower $k$-states is quite efficient due to the small energy differences. Two-dimensional spectroscopy is particular suited to study intraband relaxation since it provides directly the excitation dependence of the dynamics. Using this technique it was found that in the case of 1,1'-diethyl-3,3'-bis(sulfopropyl)-5,5',6,6'-tetrachlorobenzimidacarbocyanine aggregates the rate for intraband relaxation accelerates from zero at the red wing of the absorption band to 1/(40Êfs) in the blue wing (\cite{Sti06}).

\begin{figure}
\begin{center}
\includegraphics[width=0.8\textwidth]{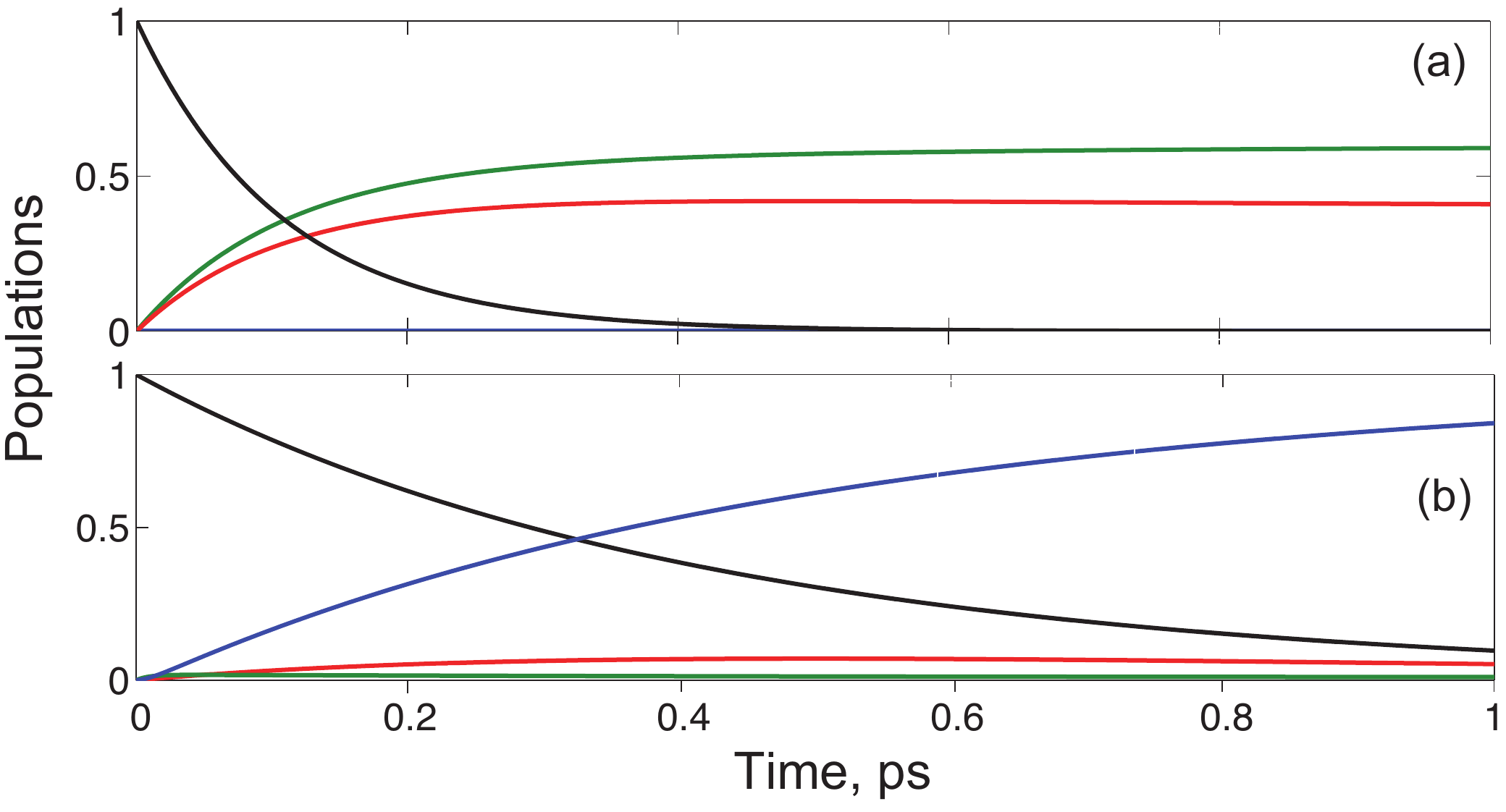}
\caption{Population relaxation within the vibronic one-exciton manifold provided that the highest state is initially prepared (cf. arrows in Fig. \ref{fig:VD_lev}). (a) $J/\omega_{\rm vib}=2$ (curves at 1~ps correspond to states $\alpha_1=4,3,2$ with increasing value). (b) $J/\omega_{\rm vib}=0.3$ (curves at 1 ps correspond to states $\alpha_1=2,3,4,1$ with increasing value). The upper panel shows the effect of transient population trapping (for more details, see \cite{polyutov11:xxx}).}
\label{fig:VD_pop}
\end{center}
\end{figure}

The intraband relaxation might be influenced by the vibronic substructure of the exciton band. As an example the population relaxation in the one-exciton band according to Redfield theory is shown in Fig. \ref{fig:VD_pop} for the vibronic dimer model of Fig. \ref{fig:VD_lev}. Most notable is the so-called transient population trapping observed in the case of strong Coulomb coupling (panel (a)). With increasing ratio $J/\omega_{\rm vib}$ the level separation increases (cf. Fig. \ref{fig:VD_lev}) such that the transition frequency to reach the lowest state moves out of the frequency range where the spectral density of the coupled bath modes differs from zero. Thus on the time scale given, no equilibrium population is established within the one-exciton vibronic manifold (\cite{polyutov11:xxx}).
Finally, we point out that the effect of vibronic coupling can also be viewed in real space, i.e. in the context of exciton transport. Here it has been reported that vibronic coupling may have a strong influence on the  exciton coherence size (\cite{roden09:044909,womick11:1347}). 

\subsection{Interband Relaxation}
The two-exciton band consists of states in which two excitons are on the coherence domain and are coupled to monomeric double excitations. Up to now the population of a two-exciton state in aggregates has not yet been unambiguously observed. This is usually attributed to the short lifetime of such a state. A blue shifted component of the transient absorption features was found during a short time interval after the excitation in the case of merocyanine J-aggregates (\cite{Ham02}) and of PIC-Br aggregates (\cite{Min94}). It was attributed to the annihilation induced decay of multiple excitons occupying one coherence domain. However, at a very low temperature of 1.5~K the width of the one- to two-exciton transition in PIC-Br aggregates is only 5~cm$^{-1}$ (\cite{Fid93}). This indicates that two-exciton states with a lifetime of at least 1~ps should exist under these conditions.

\begin{figure}
\begin{center}
\includegraphics[width=0.6\textwidth]{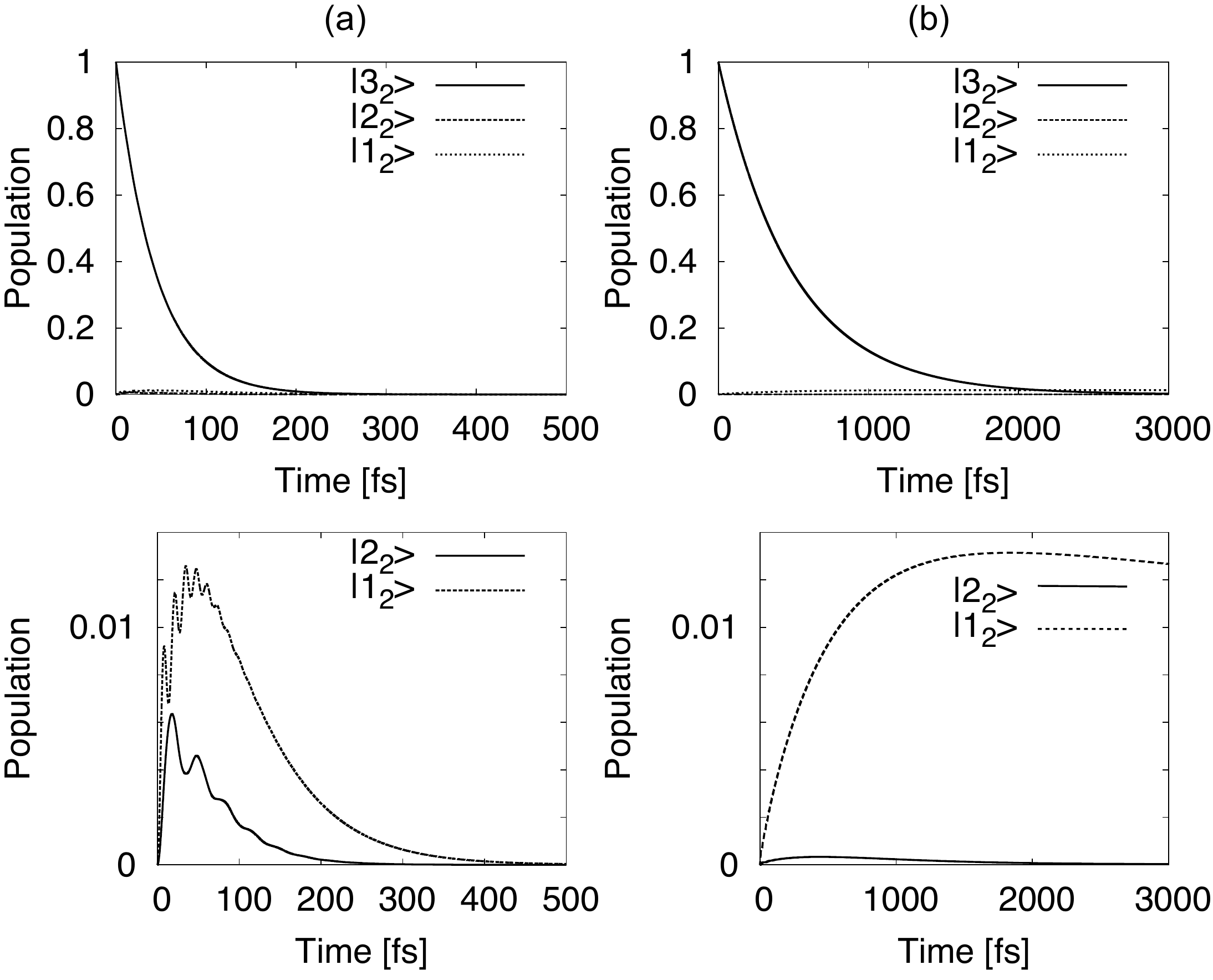}
\caption{Population dynamics of the two-exciton states of the heterodimer model of Figs. \ref{fig:tpa}a ($\Delta_n=0.2$ eV, panels (a)) and \ref{fig:tpa}c ($\kappa_n=2$, panels (b)) which follows from a hierarchy equation of motion approach including an internal conversion rate of 1/500~fs. Notice that due to intraband relaxation in the case of panel (b) there is some intermediate population trapping in state $\alpha_2=1$ which has a pronounced nonlocal character (\cite{yan11:xxx}).}
\label{fig:anni}
\end{center}
\end{figure}

The microscopic origin of the interband transition is the breakdown of the Born-Oppenheimer approximation. According to Section \ref{sec:EEA} it is the local character of the exciton states which determines the transition rate, Eq. (\ref{eq:EEA}). To demonstrate this in Fig. \ref{fig:anni} we show the population dynamics of the heterodimer model of Fig. \ref{fig:tpa} as obtained using a hierarchy equation of motion approach (\cite{yan11:xxx}). For simplicity it is assumed that the highest state of the two-exciton band is populated. Panels (a) correspond to the case of Fig. \ref{fig:tpa}a ($\Delta_n=0.2$ eV), i.e. local and nonlocal double excitations are weakly  mixed. Since the initial state has a pronounced local character the decay via interband transitions is rather rapid. In contrast in panel (b) the case of Fig. \ref{fig:tpa}c ($\kappa_n=2$) is used, where nonlocal and local states are strongly mixed. Having a more pronounced nonlocal character the initial state decays much slower.

Intraband relaxation is also very important if several excitonic J-bands exist within the same aggregate. In the case of porphyrin J-aggregates a decay time of about 1~ps was observed for the relaxation from the exciton band formed by the molecular S$_2$-states down to the band resulting from the molecular S$_1$-states (\cite{Mis99}). J-aggregates of Tetrakis(4-sulfonatophenyl)porphyrin (TPPS4) exhibit also two J-bands, a strong one around 490~nm and a weaker one at 707~nm. The exciton coherence length in the upper band is about 10 chromophores. After optical excitation of this band ultrafast decay to the lower band occurs and the coherence length reduces to one or two molecules (\cite{gulbinas07:255}). Subsequently the exciton relaxes to a not or only weekly radiating state within 200~ps in agreement with the observation that the fluorescence yield decreases with aggregation. This kind of self-trapping is not very common in J-aggregates. In TPPS4 the transition dipole to the lowest excited electronic state is small and thereby also the excitonic coupling what might favor localization of the exciton.
In complex aggregates with several overlapping J-bands two-dimensional spectroscopy is a suitable tool to disentangle various contributions to the relaxation dynamics as well as interband coherences (\cite{milota09:45}). For an aggregate with four overlapping bands a hierarchy of relaxation steps in the range of 100~fs to 1~ps was observed in this way (\cite{Nem09}).

\subsection{Exciton-Exciton Annihilation}
Laser spectroscopic experiments found that the exciton lifetime in J-aggregates decreases with excitation density and time resolved measurements revealed that the associated kinetics is non-exponential (\cite{sundstrom88:2754}). These phenomena occur already at moderate excitation densities and result from EEA. Due to their mobility excitons can approach each other and interact. In such an encounter one exciton is deactivated with a high probability while the other remains. The mechanism was already introduced in the Introduction and in Section \ref{sec:EEA}. In the simplest case the time dependent exciton density $n(t)$ obeys the rate equation, Eq. \re{eq:anni}, supplemented by a term describing the intrinsic decay with rate $k_1$.

\begin{figure}
\begin{center}
\includegraphics[width=0.5\textwidth]{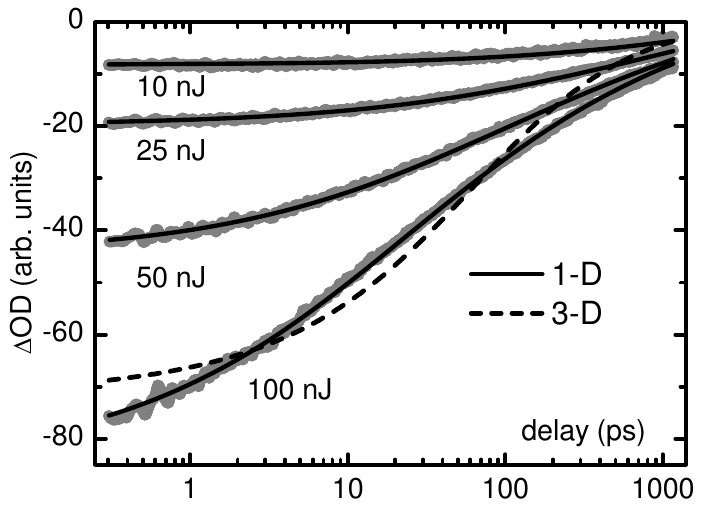}
\caption{Transient absorption signal (grey dots) of excitons on PBI J-aggregates for different excitation energies. The observed kinetics is modeled by assuming EEA and one- (black solid line) and three-dimensional diffusion (broken line) of the excitons (see also \cite{Mar11}).
}
\label{fig:anniexp}
\end{center}
\end{figure}

For a signal $S$ which is proportional to $n$ plotting $dS/dt/S$ versus the signal itself should give a linear rise. For PIC chloride a deviation from this behavior was observed at low excitation densities indicating that annihilation occurs only above a certain threshold density (\cite{sundstrom88:2754}). From this density one can estimate the size of the domain over which an exciton migrates and in which at least two excitons have to be generated to induce annihilation. In this way it was shown that in PIC chloride excitons can sample up to 10$^4$ monomers (\cite{sundstrom88:2754}). For THIATS exciton migration over $6\times 10^4$ monomers at room temperature and even over $6\times 10^6$ at 77~K was observed (\cite{Sche98}).

A more sophisticated analysis was carried out for EEA on PBI J-aggregates (\cite{Mar11}). Figure \ref{fig:anniexp} shows the time dependent absorption signal due to optical excitation of PBI aggregates in methylcyclohexane for different excitation energies. The negative signal is proportional to the number of excitons and is obtained by integrating the spectral region from 590~nm to 625~nm, where bleach and stimulated emission are dominating the transient absorption spectrum (cf. Fig. \ref{fig:TA}). The exciton signal exhibits a non-exponential decay which accelerates with increasing excitation density indicating that EEA takes place. The time-dependent exciton concentration is modeled taking annihilation into account and assuming that the excitons can diffuse within the aggregates. In this case $\gamma (t)$ of the above rate equation becomes time dependent and adopts the form $\gamma _{\rm 1D} (t)=\sqrt{2D/\pi t}$ for one-dimensional and $\gamma _{\rm 3D} (t)=4\pi RD(1+R/\sqrt{2\pi Dt})$ for three-dimensional diffusion. $D$ is the diffusion constant and $R$ the critical distance between two excitons when annihilation takes place. For the three-dimensional case, insufficient agreement with the data is achieved while the one-dimensional model reproduces the measurements almost perfectly (see Fig. \ref{fig:anniexp}). This indicates that the excitons move along the PBI aggregates as on one-dimensional strings. The analysis reveals also a diffusion constant of 1.29~nm$^2$/ps (\cite{Mar11}). It demonstrates that annihilation dynamics can be used to investigate local transport properties.
%
\section{Concluding Remarks}
%
Excitation energy transfer in molecular aggregates has a long history, but only recent years have witnessed an unprecedented step forward in our mechanistic understanding of the underlying photophysical properties. Ultrafast nonlinear spectroscopy provides insight into the real time dynamics of exciton motion, unravelling its coherences and their decoherence due to the interaction between excitonic and nuclear degrees of freedom. For example, in the past the concept of an exciton coherence domain has been inferred from indirect indicators such as band shapes or emission lifetimes. Nowadays two-dimensional spectroscopy provides a much more direct look and is capable to disentangle time scales and correlations of fluctuations. This development comes along with a much increased prospect for realistic modeling. Still the Frenkel exciton Hamiltonian and models like the dimer or the linear chain are indispensable. However, the development in Computational Chemistry allows for the first time to obtain first principles based parametrizations for concrete systems.
Last but not least, organic synthesis has taken up the challenge to design and synthesize  dye aggregates with desired electronic properties. The convergence of these different areas holds great promises for the future development, which includes, for instance,  the integration of molecular aggregates into organic solar cell devices.

%
\section*{Acknowledgments}
We gratefully acknowledge financial support by the Deutsche Forschungsgemeinschaft (Sfb 652).
%


\end{document}